\documentclass[twocolumn]{aastex62}

\usepackage{booktabs}

\accepted{May 26, 2020}

\submitjournal{ApJ}

\shorttitle{A White Dwarf with Transiting Circumstellar Material}
\shortauthors{Vanderbosch et al.}

\begin{document}
	
\title{A White Dwarf with Transiting Circumstellar Material Far Outside the Roche Limit}

\author{Z. Vanderbosch}
\affil{Department of Astronomy, University of Texas at Austin, Austin, TX-78712, USA}
\affil{McDonald Observatory, Fort Davis, TX-79734, USA}
\email{zvanderbosch@astro.as.utexas.edu}

\author{J.~J. Hermes}
\affil{Department of Astronomy, Boston University, Boston, MA-02215, USA}

\author{E. Dennihy}
\affil{Gemini Observatory, Casilla 603, La Serena, Chile}

\author{B.~H. Dunlap}
\affil{Department of Astronomy, University of Texas at Austin, Austin, TX-78712, USA}

\author{P. Izquierdo}
\affil{Instituto de Astrof\'isica de Canarias, 38205 La Laguna, Tenerife, Spain}
\affil{Departamento de Astrof\'isica, Universidad de La Laguna, 38206 La Laguna, Tenerife, Spain}

\author{P.-E. Tremblay}
\affil{Department of Physics, University of Warwick, Coventry CV4 7AL, UK}

\author{P.~B. Cho}
\affil{Department of Astronomy, University of Texas at Austin, Austin, TX-78712, USA}
\affil{McDonald Observatory, Fort Davis, TX-79734, USA}

\author{B.~T. G\"ansicke}
\affil{Department of Physics, University of Warwick, Coventry CV4 7AL, UK}

\author{O. Toloza}
\affil{Department of Physics, University of Warwick, Coventry CV4 7AL, UK}

\author{K.~J. Bell}
\affil{DIRAC Institute, Department of Astronomy, University of Washington, Seattle, WA-98195, USA}
\affil{NSF Astronomy and Astrophysics Postdoctoral Fellow}

\author{M.~H. Montgomery}
\affil{Department of Astronomy, University of Texas at Austin, Austin, TX-78712, USA}
\affil{McDonald Observatory, Fort Davis, TX-79734, USA}

\author{D.~E. Winget}
\affil{Department of Astronomy, University of Texas at Austin, Austin, TX-78712, USA}
\affil{McDonald Observatory, Fort Davis, TX-79734, USA}


\begin{abstract}
	We report the discovery of a white dwarf exhibiting deep, irregularly shaped transits, indicative of circumstellar planetary debris. Using Zwicky Transient Facility DR2 photometry of ZTF\,J013906.17+524536.89 and follow-up observations from the Las Cumbres Observatory, we identify multiple transit events that recur every ${\approx}\,107.2$\,d, much longer than the 4.5--4.9\,h orbital periods observed in WD\,1145+017, the only other white dwarf known with transiting planetary debris. The transits vary in both depth and duration, lasting 15--25\,d and reaching 20--45\,\% dips in flux. Optical spectra reveal strong Balmer lines, identifying the white dwarf as a DA with $T_{\mathrm{eff}}=10{,}530\pm140\,\mathrm{K}$ and $\log(g)=7.86\pm0.06$. A Ca\,{\sc ii}~K absorption feature is present in all spectra both in and out of transit. Spectra obtained during one night at roughly 15\,\% transit depth show increased Ca\,{\sc ii}~K absorption with a model atmospheric fit suggesting $[\mathrm{Ca/H}]=-4.6\pm0.3$, whereas spectra taken on three nights out of transit have $[\mathrm{Ca/H}]$ of -5.5, -5.3, and -4.9 with similar uncertainties. While the Ca\,{\sc ii}~K line strength varies by only 2-sigma, we consider a predominantly interstellar origin for Ca absorption unlikely. We suggest a larger column density of circumstellar metallic gas along the line of site or increased accretion of material onto the white dwarf's surface are responsible for the Ca absorption, but further spectroscopic studies are required. In addition, high-speed time series photometry out of transit reveals variability with periods of 900 and 1030\,s, consistent with ZZ Ceti pulsations.
	
\end{abstract}

\keywords{White dwarf stars --- Eclipses --- Transits --- Debris disks --- Circumstellar dust --- Circumstellar gas --- Roche limit --- Tidal disruption --- Stellar pulsations --- Planetesimals}

\section{Introduction} \label{sec:intro}

The vast majority of currently known planet hosts will one day become white dwarfs, the end products of stellar evolution for low- to intermediate-mass stars ($M<10\,M_{\odot}$, \citealt{Williams_2009_1}). Many planets are expected to survive the post-main-sequence evolution of their host stars \citep{Veras_2016_1}. Indeed, at least one third of all known white dwarfs with $T_{\mathrm{eff}}<20{,}000\:\rm{K}$ exhibit heavy elements beyond hydrogen and helium in their photospheres \citep{Zuckerman2010, Koester2014}, which is commonly interpreted as the active accretion of tidally disrupted planetary debris \citep{Debes2002,Jura2003,Zuckerman2010,Veras2014,Farihi_2016_1,Mustill_2018_1}.

\begin{figure*}[t!]
	\epsscale{1.18}
	\plotone{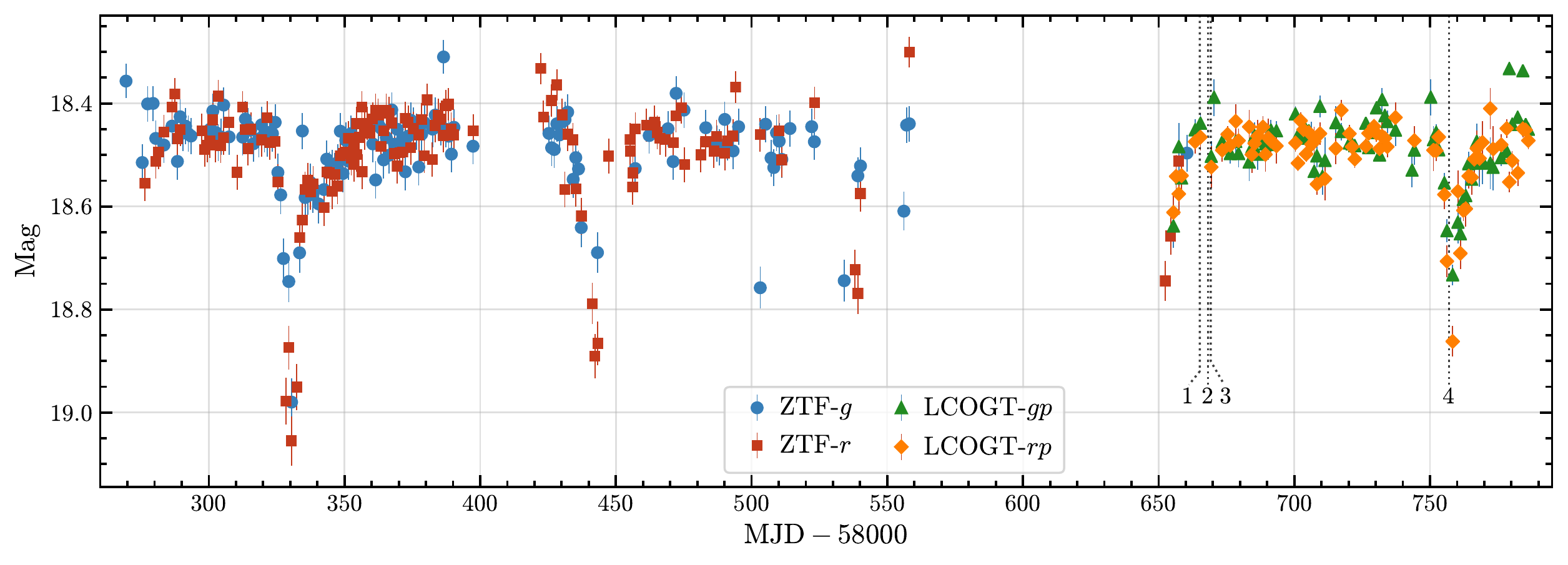}
	\caption{The light curve for ZTF\,J0139+5245 from ZTF DR2 and LCOGT observations. Three full transits are observed along with a partial transit at the start of LCOGT monitoring ($\mathrm{MJD}\,{\simeq}\,58650$). While sparsely observed, an additional transit event may be seen near $\mathrm{MJD}\,{\simeq}\,58540$. Magnitudes in the ZTF-\textit{g} and \textit{r} bands are shown with blue circles and red squares, respectively, while magnitudes in the LCOGT-\textit{gp} and \textit{rp} bands and shown with green triangles and orange squares, respectively. The average MJD for each of the four nights of WHT/ISIS spectroscopic observations is labeled with a vertical dotted line. \label{fig:phot}}
\end{figure*}

This debris has been detected in more than 40 white dwarfs as an infrared excess indicative of circumstellar dust, while in a small number of these systems, a gaseous debris component is detected via Ca\,{\sc ii} triplet emission \citep{Farihi_2016_1}. Only one white dwarf, however, exhibits transits caused by planetary material, WD\,1145+017 \citep{Vanderburg2015}, which displays both photometric and spectroscopic signatures of transiting debris in $4.5{-}4.9$\,h orbits.

In this manuscript, we report the discovery of a second white dwarf, ZTF\,J013906.17+524536.89 (hereafter ZTF\,J0139+5245), exhibiting photometric and spectroscopic evidence for transits caused by circumstellar planetary debris. In the following sections, we present publicly archived and newly obtained photometry and spectroscopy that constrain the white dwarf and its circumstellar material.

\section{Observations} \label{sec:observations}

\subsection{Public ZTF Photometry}

\defcitealias{GF2019}{GF19}

We discovered two transits in ZTF\,J0139+5245 during a general search for variable white dwarfs in the public Zwicky Transient Facility (ZTF) survey \citep{Masci2019,Bellm2019} by cross-matching the {\em Gaia} DR2 catalogue of white dwarfs \citep[hereafter \citetalias{GF2019}]{GF2019} with the public ZTF transient alert database \citep{Patterson2019} using the API provided by the Las Cumbres Observatory Make Alerts Really Simple (MARS) project\footnote{\url{https://mars.lco.global/}}. Prior to the cross-match, we trimmed the full \citetalias{GF2019} catalogue into an astrometrically clean, 200\,pc sample of ${\approx}\,40{,}000$ objects using the criteria recommended by \citet{Lindegren2018} and \citet{Evans2018}. As of 2019 May 08, this cross match results in 783 objects that have at least one alert from ZTF indicative of transient or periodically variable behavior. We downloaded the public ZTF DR2 light curves for each object with an alert and visually inspected them and their periodograms for signs of variability. ZTF\,J0139+5245 stood out as the only object with long-lasting, well-defined dips in flux (see Fig.~\ref{fig:phot}).

In total, ZTF\,J0139+5245 was covered by 279 public ZTF observations deemed of good photometric quality by the ZTF pipeline.  We filtered these observations by only selecting points where \textit{catflags}$\:=0$, a condition for generating clean light curves recommended in the ZTF Science Data System Explanatory Supplement\footnote{\url{http://web.ipac.caltech.edu/staff/fmasci/ztf/ztf_pipelines_deliverables.pdf}}. The final light curve for ZTF\,J0139+5245, shown in Figure~\ref{fig:phot}, has 266 data points (128 in \textit{g} and 138 in \textit{r}) with a median temporal separation between observations of 22.4\,h with occasional large, multi-day gaps. ZTF photometry is on the AB magnitude scale, calibrated using Pan-STARRS1 Survey (PS1) sources \citep{Masci2019}.

\subsection{LCOGT Photometry}

We began monitoring ZTF\,J0139+5245 on 2019 June 21 using the Las Cumbres Observatory Global Telescope (LCOGT) 1.0-m telescope network and have since observed the third full transit along with a partial transit event (see Fig.~\ref{fig:phot}).

We requested one to six observations each night with exposures between two and three\,min long in both \textit{gp} and \textit{rp} filters using the Sinistro imaging instrument \citep{Brown_2013_1}. Completed observations were bias, dark, and flat-field corrected via the {\sc banzai} pipeline\footnote{\url{https://github.com/LCOGT/banzai}}. We performed circular aperture photometry on all sources detected within each image using the {\tt Photutils} package in Python \citep{Bradley_2019_2533376} and then cross-matched them with known PS1 sources. We measured the difference between PS1 and instrumental magnitudes ($m_d=m_{\mathrm{PS1}}-m_{\mathrm{LCOGT}}$) while filtering for outliers and likely galaxy candidates. We converted our instrumental magnitudes to apparent magnitudes on the PS1 scale by solving for a zero-point offset ($z$) and color term ($c$) with a least squares fit to $m_d=z + c(g_{\mathrm{PS1}}-r_{\mathrm{PS1}})$.

\subsection{WHT/ISIS Spectroscopy}

\begin{figure}[t]
	\epsscale{1.15}
	\plotone{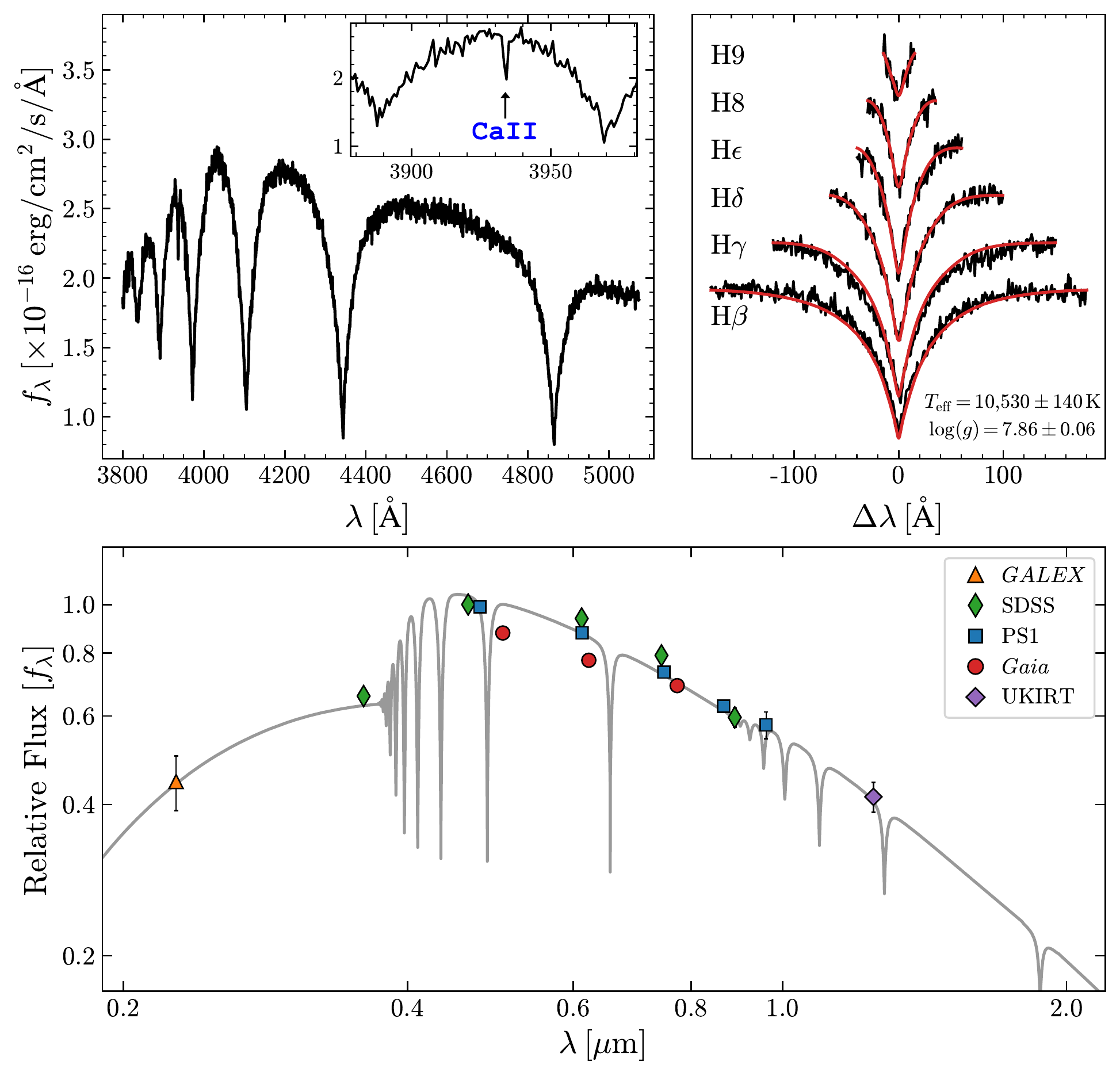}
	\caption{The WHT/ISIS optical spectrum and SED of ZTF\,J0139+5245. The top left panel shows the combined spectrum for the first three nights of observations, taken out-of-transit, with an inset plot highlighting the Ca\,{\sc ii} K absorption feature at 3934$\mathrm{\AA}$. The top right panel shows the fit to the six Balmer lines, $\mathrm{H}\beta-\mathrm{H}9$, from which we derive spectroscopic $\log(g)$ and $T_{\mathrm{eff}}$ values (see Section \ref{sec:starparams}). The bottom panel shows the SED with photometric data de-reddened using the reddening law of \citet{Fitzpatrick1999} and \citet{Indebetouw2005} with $E(B\,{-}\,V)=0.12$ and $A_V=0.38$. We over-plot a model spectrum \citep[grey line,][]{Koester_2010_1} with $T_{\mathrm{eff}}=10{,}500\,\mathrm{K}$ and $\log(g)=7.75$ representing a good match to the shape of the observed SED.  \label{fig:spectrum}}
\end{figure}

We carried out spectroscopic observations of ZTF\-J0139+5245 on 2019 June 30, July 3, July 4, and September 30 using the Intermediate-dispersion Spectrograph and Imaging System (ISIS), mounted on the 4.2-m William Herschel Telescope (WHT) in La Palma, Spain (see Figs.~\ref{fig:spectrum} \& \ref{fig:ca_spec}). The first three nights occurred outside of any deep transit event, but did occur just a few days after the nearest transit. Limits on transit depth for these three nights are $<5$\%. The fourth night occurred during a transit event at roughly 15\% transit depth (see labels 1-4 in Figs.~\ref{fig:phot} \& \ref{fig:ca_spec}).

We performed the observations using a $1''$ slit and 600 line $\mathrm{mm}^{-1}$ grating in the blue ($3800-5200\,\mathrm{\AA}$) arm of the spectrograph, achieving 1.9~\AA\ resolution. Each night we took a series of consecutive 20-min exposures under clear sky conditions with seeing values of $0.6''-1.0''$ and an average airmass of 1.5. We obtained a total of 17 exposures on nights one through three, and six exposures on night four, for combined exposure times of 5.7\,h and 2.0\,h, respectively.

We applied bias, flat-field, and cosmic ray corrections to our spectra using standard procedures within {\sc iraf}. We optimally extracted the one-dimensional spectrum \citep{Horne1986-1} using the data reduction software {\label{pamela}\sc pamela}.  We used {\sc molly} \citep{Marsh1989} to wavelength and flux-calibrate the spectra by fitting a fourth-order polynomial to the HgArNeXe arc data and a five-knot spline to the spectrophotometric standard star, respectively. Arcs and standards were obtained using the same instrument setup just before and after the science observations were carried out.

\subsection{Additional Survey Data} \label{sec:SED}

We compiled all available photometric and astrometric data for ZTF\,J0139+5245 from the {\em Galaxy Evolution Explorer} ({\em GALEX}, \citealt{Martin2005,Morrissey2005}), {\em Gaia} DR2 \citep{Gaia2}, the Sloan Digital Sky Survey (SDSS) DR9 \citep{Ahn2012}, the Pan-STARRS1 Survey \citep{Chambers2016} DR2, and the United Kingdom Infra-Red Telescope (UKIRT) Hemisphere Survey (UHS, \citealt{Dye_2019_1}). The compiled data are summarized in Table \ref{tab:summary} while the spectral energy distribution (SED) is displayed in Fig. \ref{fig:spectrum}.

\begin{table}[t!]
\renewcommand{\thetable}{\arabic{table}}
\centering
\caption{ZTF\,J0139+5245 Summary of Properties} \label{tab:summary}
\begin{tabular}{rll}
\tablewidth{0pt}
\hline
\hline
$\alpha$, $\delta$ (J2000)  & \multicolumn{2}{l}{$01^{\mathrm{h}}39^{\mathrm{m}}06^{\mathrm{s}}.17$, ${+}52^{\circ}45'36''.89^{[1]}$} \\
$\mu_{\alpha}$, $\mu_{\delta}$ (mas/yr) & \multicolumn{2}{l}{$87.32\pm0.39$, $4.99\pm0.43^{[1]}$} \\
$\varpi$ (mas)              & $5.77\pm0.25^{[1]}$       & \\
$d$ (pc)                    & $172.9\pm7.4^{[1]}$       & \\
$T_{\mathrm{eff}}^{3\mathrm{D}}$ (K)& \multicolumn{2}{l}{$10{,}530\pm140^{[2]}$} \\
$\log(g)^{3\rm{D}}$ (cgs)   & \multicolumn{2}{l}{$7.86\pm0.06^{[2]}$}    \\
$M_*$ ($M_{\odot}$)         & $0.52\pm0.03^{[2]}$       & \\
{\sc nuv}                   & $19.89\pm0.14^{[3,4]}$    & \\
$G$                         & \multicolumn{2}{l}{$18.594\pm0.008^{[1]}$} \\
$G_{\rm{BP}}$               & $18.55\pm0.02^{[1]}$      & \\
$G_{\rm{RP}}$               & $18.64\pm0.03^{[1]}$      & \\
\,                          & {\sc sdss}$^{[5]}$        & {\sc ps}{\scriptsize 1}$^{[6]}$ \\
\cmidrule(lr){2-3}
$u$                         & $19.04\pm0.04$            & \\
$g$                         & $18.46\pm0.01$            & $18.48\pm0.01$ \\
$r$                         & $18.40\pm0.01$            & $18.49\pm0.01$ \\
$i$                         & $18.52\pm0.01$            & $18.59\pm0.01$ \\
$z$                         & $18.75\pm0.04$            & $18.69\pm0.02$ \\
$y$                         &                           & $18.78\pm0.03$ \\
\cmidrule(lr){2-3}
$J$                         & $19.10\pm0.07^{[7]}$      & \\
\hline
\multicolumn{3}{l}{\footnotesize{{\sc Note} -- All magnitudes are on the {\sc ab} scale. {\em Gaia} and {\sc ukirt}}} \\ [-0.08cm]
\multicolumn{3}{l}{\footnotesize{were converted from Vega to {\sc ab} scales using the {\em Gaia} {\sc dr}{\scriptsize 2}}} \\ [-0.1cm]
\multicolumn{3}{l}{\footnotesize{documentation\footnote{\url{https://gea.esac.esa.int/archive/documentation/GDR2/}} and \citet{Hewett_2006_1}, respectively.}} \\ [-0.1cm]

\multicolumn{3}{l}{\footnotesize{[1] \citet{Gaia2}, [2] This Work,}} \\ [-0.1cm]
\multicolumn{3}{l}{\footnotesize{[3] \citet{Martin2005}, [4] \citet{Morrissey2005},}} \\ [-0.1cm]
\multicolumn{3}{l}{\footnotesize{[5] \citet{Ahn2012}, [6] \citet{Chambers2016},}} \\ [-0.1cm]
\multicolumn{3}{l}{\footnotesize{[7] \citet{Dye_2019_1}.}} \\ [-0.1cm]

\end{tabular}
\end{table}

The PS1 $3\pi$ survey \citep{Kaiser2010,Magnier2013} is multi-epoch, so we also obtained all individual detections to look for additional transit events. In total, ZTF\,J0139+5245 was observed by PS1 63 times over the course of 4.4\,yr. To filter out poor-quality detections, we first removed those where $>5\%$ of the fitted PSF model was contaminated by bad pixels. To further filter our data, we followed the methods of \citet{Fulton2014}. The resulting PS1 data contain 52 epochal detections (\textit{g}:11, \textit{r}:9, \textit{i}:11, \textit{z}:10, \textit{y}:11) with 30--80\,s exposure times.

\subsection{McDonald 2.1-m Photometry}

We acquired high-speed time-series photometry on four nights between 2019 June 26 and July 1 and on five nights between 2019 August 27 and September 3 using the Princeton Instruments ProEM frame-transfer CCD attached to the McDonald Observatory 2.1-m Otto Struve telescope (see Fig.~\ref{fig:mcdonald}). We used Astrodon Gen2 Sloan $g'$ and $r'$ filters in an automated filter wheel with 15 to 30-s exposure times. Target availability limited our June-July runs to ${\sim}1\,$h each night, while our August-September runs ranged between 2.6 and 5.3\,h, for a total of $23.1$\,h. We used {\sc iraf} to bias, dark, and flat-field correct the McDonald data using standard calibration frames taken before each night of observations. We performed circular aperture photometry using the {\sc iraf} routine {\sc ccd\_hsp} \citep{Kanaan2002}. Lastly, we used the {\sc Wqed} software suite to generate light curves with the optimal aperture size and apply a barycentric correction to the mid-exposure timestamp of each image \citep{Thompson2013}.

\section{Results} \label{sec:stellar}

\subsection{White Dwarf Atmospheric Parameters} \label{sec:starparams}

Prior to obtaining a spectrum, ZTF\,J0139+5245 was considered a high probability white dwarf candidate due to both its SDSS photometric colors \citep{Girven_2011_1} and its location in the {\em Gaia} color magnitude diagram \citepalias{GF2019}. Utilizing {\em Gaia} photometry and parallax, \citetalias{GF2019} report $T_{\mathrm{eff}}=9{,}420\pm580\:\mathrm{K}$, $\log(g)=7.87\pm0.21$, and $M_*=0.52\pm0.11\:M_{\odot}$ for ZTF\,J0139+5245 assuming a pure-H atmosphere.

The newly obtained spectra provide clear evidence that ZTF\,J0139+5245 is a white dwarf of spectral type DA due to the presence of broad H-Balmer absorption features (see Fig.~\ref{fig:spectrum}). Using the combined spectrum for the three nights of observations taken out-of-transit, we fit six Balmer lines, $\rm{H}\beta - \rm{H}9$, utilizing the one-dimensional (1D) models and fitting procedures described in \citet{Tremblay2011}. We find $T_{\mathrm{eff}}^{1\mathrm{D}}\,{=}\,10{,}790\pm140\,\mathrm{K}$ and $\log(g)^{1\rm{D}}=8.09\pm0.06$ whose formal uncertainties have been added in quadrature to $1.2\,\%$ $T_{\mathrm{eff}}$ and 0.038-dex $\log(g)$ uncertainties to account for typical systematics \citep{Liebert2005}. We apply corrections to these 1D values based on the three-dimensional (3D) convection simulations of \citet{Tremblay2013} to obtain $T_{\mathrm{eff}}^{3\mathrm{D}}\,{=}\,10{,}530\pm140\,\mathrm{K}$ and $\log(g)^{3\mathrm{D}}\,{=}\,7.86\pm0.06$.  Utilizing the 3D values and the white dwarf evolutionary models of \citet{Fontaine2001} with evenly mixed C/O cores and thick-H layers, we obtain $M_*=0.52\pm0.03\,M_{\odot}$.

Using the best fit spectroscopic temperature, we find a good match between a single white dwarf model SED \citep{Koester_2010_1} and the observed SED with extinction and reddening corrections applied to the observed SED of $A_V\,{=}\,0.38$ and $E{(B{-}V)}\,{=}\,0.12$ (see Fig.~\ref{fig:spectrum}). The distance scaling relation of \citetalias{GF2019} estimates less extinction for ZTF\,J0139+5245 with $A_V\,{=}\,0.11$, as does the 3D reddening map of \citet{Capitanio_2017_1} with $A_V\,{=}\,0.16\pm0.06$. The full Galactic extinction along the line of sight at ZTF\,J0139+5245's coordinates is large, however (0.84, \citealt{Schlafly2011}), and may be underestimated at its relatively nearby distance ($d=172.9\pm7.4\,\mathrm{pc}$, \citealt{Bailer-Jones2018}) due to locally dense regions within the ISM. The additional extinction may also be caused by the circumstellar material in this system.

\begin{figure}[t]
	\epsscale{1.18}
	\plotone{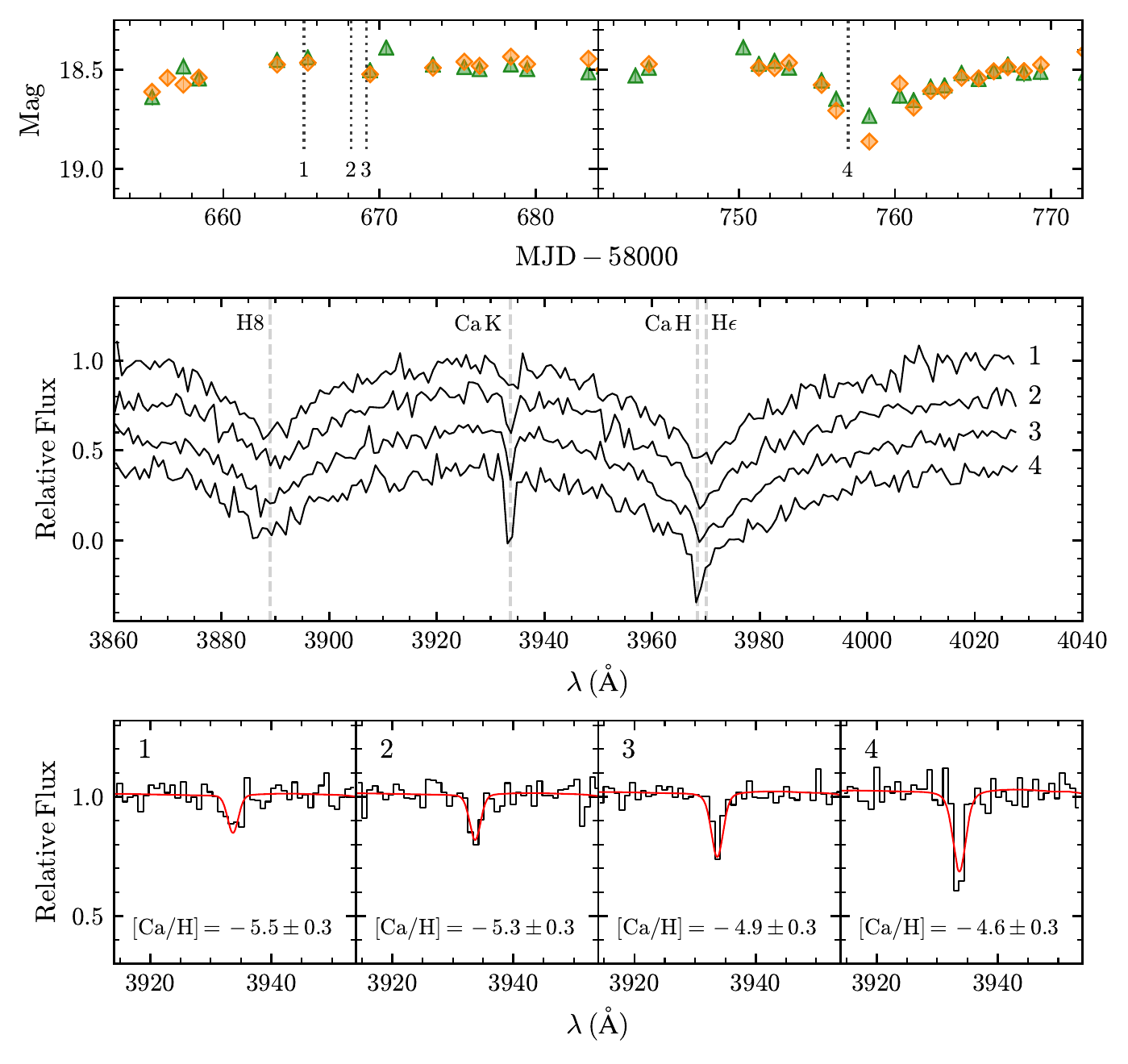}
	\caption{The WHT/ISIS spectra in normalized flux units for each night, focused on the Ca\,{\sc ii} H \& K absorption features. The top panels show zoomed in portions of the LCOGT \textit{gp} (blue circles) and \textit{rp} (red squares) photometry with the average MJD for each night's combined spectrum labeled with a vertical dotted line. The middle panel shows a broader wavelength region with spectra vertically stacked for comparison, while the bottom panels display model fits (red line) to the Ca\,{\sc ii} K line assuming a purely photospheric origin for Ca. The derived [Ca/H] photospheric abundance is given for each night, but we caution that the observed Ca absorption may also result from an increase in the column density of circumstellar metallic gas along the line of sight, in which case these [Ca/H] values are overestimates of the photospheric abundance (see Sec \ref{sec:calcium}). \label{fig:ca_spec}}
\end{figure}

\begin{figure*}[t]
	\epsscale{1.10}
	\plotone{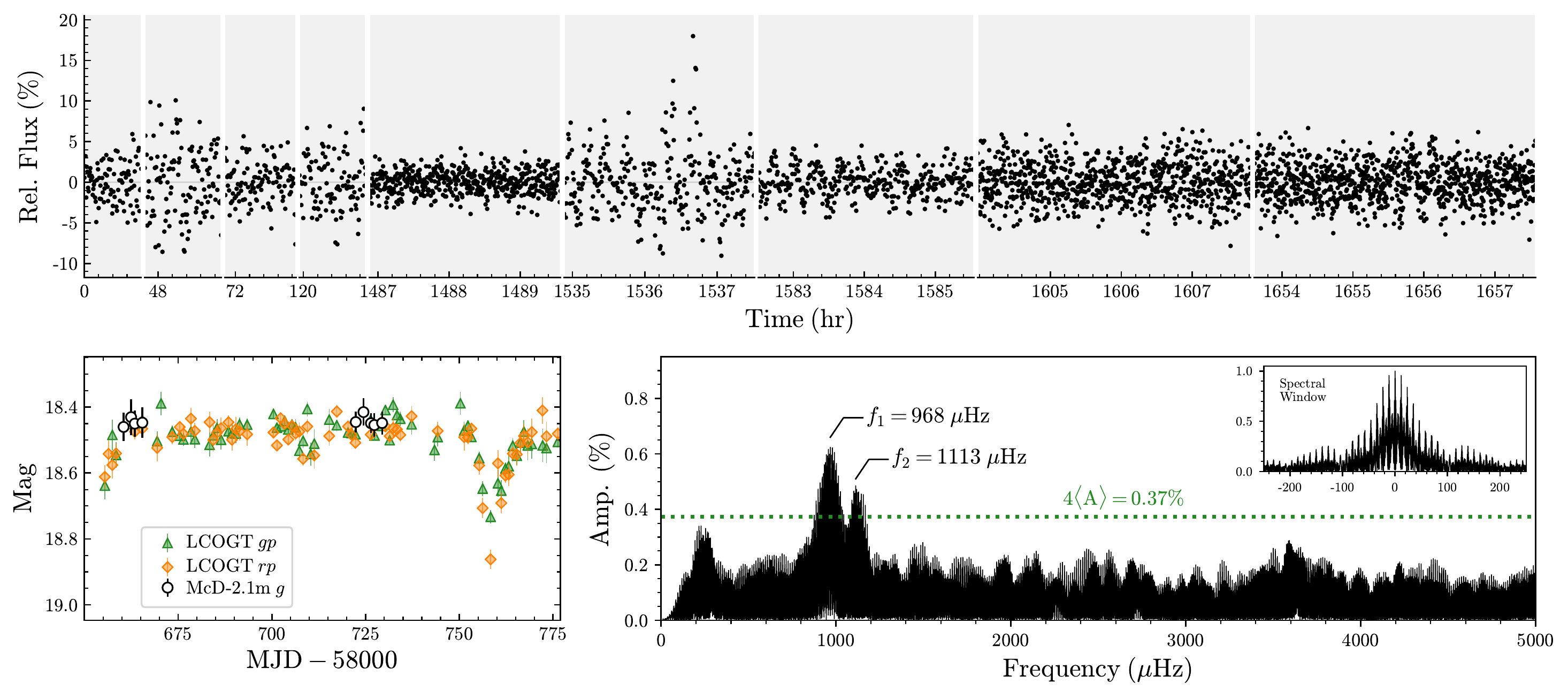}
	\caption{Nine nights of McDonald 2.1-m high speed photometry taken out of transit show variability indicative of ZZ Ceti pulsations. The top panel shows the \textit{g}- and \textit{r}-band photometry for each night, normalized relative to six comparison stars within the field of view. The bottom left panel shows the average brightness for each night of McDonald \textit{g}-band data relative to LCOGT observations, indicating the observations were taken out of transit. The Lomb Scargle periodogram of the combined light curve is shown in the bottom right panel with two significant peaks labeled by their frequencies. The periodogram's $4\langle \mathrm{A}\rangle$ significance threshold is denoted by the green dotted line. \label{fig:mcdonald}}
\end{figure*}

\subsection{Calcium Absorption} \label{sec:calcium}

In the combined spectrum for each night we detect a Ca\,{\sc ii}~K absorption feature (3934\,\AA), and on some nights a Ca\,{\sc ii}~H absorption feature as well (3968\,\AA, Fig.~\ref{fig:ca_spec}). Three possible origins can explain the detected Ca absorption: (1) gas in the interstellar medium, (2) Ca in the photosphere of the white dwarf, or (3) circumstellar metallic gas. All three may contribute to some degree.

To explore the relative importance of each possibility, we first assess the Ca\,{\sc ii}~K line strength for each night by assuming a purely photospheric origin and modeling the line using the atmosphere code of \citet{Koester_2010_1}. On nights one through three, outside of a deep transit but just a few days after the nearest transit event, we find $[\mathrm{Ca/H}]$ abundances of $-5.5\pm0.3$, $-5.3\pm0.3$, and $-4.9\pm0.3$, respectively. The uncertainties were empirically determined assuming that these first three nights are random samplings of a constant photospheric Ca abundance. On night four, at roughly 15\,\% transit depth, we find $[\mathrm{Ca/H}]\,{=}\,{-}4.6\pm0.3$ (Fig.~\ref{fig:ca_spec}). While only 2-sigma higher relative to the lowest measured abundance, the variable line strength rules against a predominantly interstellar origin.

If the observed Ca absorption is purely photospheric, these abundances would be at the upper end of Ca pollution detected in DA white dwarfs of similar temperatures \citep[see Fig.1 in][]{Koester_2006_1}, and would indicate a high rate of metal accretion onto the white dwarf whose Ca diffusion timescale is 1--10 years \citep{Koester_2006_1,Cunningham_2019_1}. There may also be contributions from circumstellar metallic gas along the line of sight, however, in which case the derived $[\mathrm{Ca/H}]$ values should be considered overestimates of the photospheric calcium abundance. 

Absorption features with both photospheric and circumstellar components would likely exhibit two features at different velocities due to the gravitational redshift at the white dwarf surface. This difference would be ${\approx}\,24\,\mathrm{km\,s^{-1}}$ for a $0.52\,M_{\odot}$ white dwarf, which corresponds to a wavelength difference of $0.3\,\mathrm{\AA}$ for Ca\,{\sc ii}~K. With a resolution of $1.9\,\mathrm{\AA}$, our spectra would be unable to resolve these components. While unresolved, if the increase in Ca absorption is due primarily to circumstellar gas with a fixed photospheric component, a small shift in the velocity of the blended line is expected. Unfortunately, we find that the velocity uncertainties for our spectra are of order $50\,\mathrm{km\,s^{-1}}$ on most nights, preventing any meaningful comparison between nightly line velocities.

We note, however, that during the observations in transit on night four, the Ca\,{\sc ii}~K line cannot be fit adequately with a purely photospheric model. For the large Ca abundance required to match the strength of the observed line, intrinsic broadening exceeds the spectral resolution of our data, whereas the observed Ca\,{\sc ii}~K line remains unresolved (Fig.~\ref{fig:ca_spec}). This suggests that the increased Ca\,{\sc ii} absorption on night four is caused partly by an increased column density of metallic gas along the line of sight. With only one night of in-transit spectroscopy, however, the exact relationship between transit events and Ca\,{\sc ii} absorption in ZTF\,J0139+5245 remains unclear. Spectroscopic observations at higher resolution across the full orbital period are needed to assess the presence of both photospheric and circumstellar components and understand their relationship with the transiting material.

\subsection{ZZ Ceti Pulsations}

The 3D spectroscopic $T_{\mathrm{eff}}$ and $\log(g)$ place ZTF\-J0139+5245 inside the ZZ Ceti instability strip \citep{Gianninas2015} where H-atmosphere white dwarfs pulsate. To assess whether any significant variability exists at periods expected of ZZ Cetis, we analyze the nine nights of McDonald 2.1-m high-speed photometry taken out of transit. For the Lomb-Scargle periodogram of the combined McDonald light curve, we estimate a 0.1\,\% false alarm probability threshold of $0.37\,\%$ using four times the average amplitude of the periodogram, $4\langle \mathrm{A}\rangle$, between 500 and 10{,}000\,$\mu\mathrm{Hz}$ \citep{Breger_1993_1,Kuschnig_1997_1}. We find two significant peaks that rise above this threshold at $968\,\mu\rm{Hz}$ and $1113\,\mu\rm{Hz}$ (see Fig.~\ref{fig:mcdonald}). 

Owing to strong aliasing caused by large gaps in the data, we assume an extrinsic error for our frequencies equal to the daily alias ($11.6\,\mu\mathrm{Hz}$), which matches the spacing between large adjacent peaks in the spectral window. The two significant peaks have periods of 1030 and 900\,s, respectively, consistent with red-edge ZZ Cetis \citep{Mukadam_2006_1}. The pulsational variability can also be seen by eye in the McDonald photometry, most notably in the sixth and seventh panels of Fig. \ref{fig:mcdonald}. The lack of visible variations on some nights is typically inferred as the destructive beating between independent modes, a common characteristic of multi-periodic ZZ Cetis, but may also be due to increased statistical noise on nights with relatively poor weather.

The presence of pulsations in ZTF\,J0139+5245 pose an additional challenge for measuring Ca line strength variations since the effective temperature of a white dwarf can change by hundreds to thousands of degrees throughout a pulsation cycle. We expect this effect to be relatively small in our analysis because the combined spectra for each night are averaged over several pulsations cycles, and the pulsation amplitudes observed appear small on average relative to other ZZ Cetis. G29-38, for instance, is both a metal polluted and high amplitude ZZ Ceti whose Ca\,{\sc ii}~K equivalent width varies by $<10$\,\% throughout a typical pulsation cycle \citep{Thompson_2010_1,Debes_2008_1,vonHippel_2007_1}. When averaged over many pulsation cycles, the equivalent width variations due to pulsations in G29-38 are expected to drop well below the 5\,\% level even in the worst case scenario with its highest amplitude pulsations \citep{vonHippel_2007_1}. In our case, even a 5\,\% effect would be small relative to the measurement uncertainties, but future studies of ZTF,J0139+5245 at high resolution and high signal-to-noise might need to take this effect into account.

\subsection{The Transit Spacing} \label{subsec:period}

\begin{figure}[t]
	\epsscale{1.15}
	\plotone{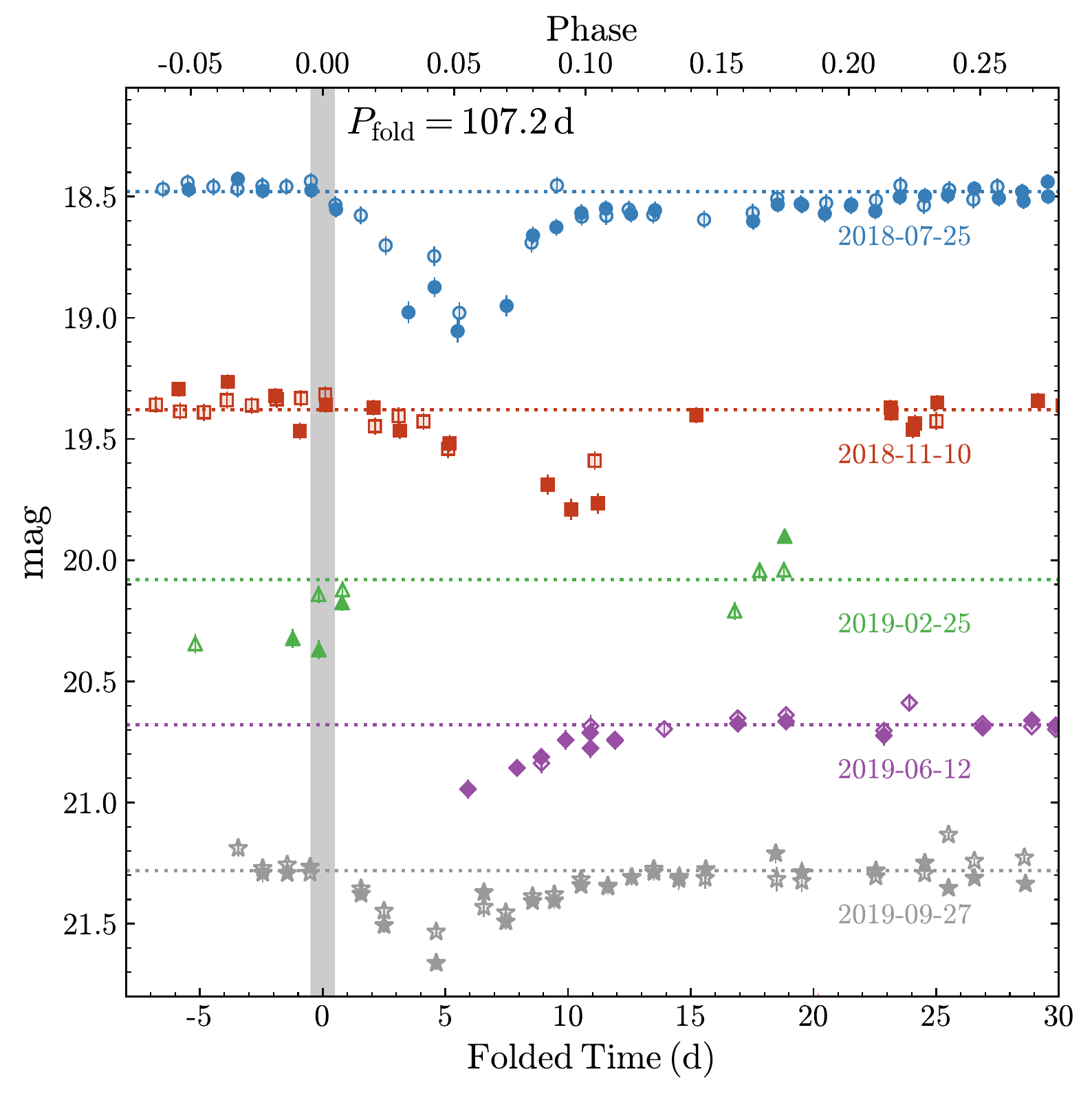}
	\caption{The ZTF and LCOGT photometry folded on a period of 107.2 days and vertically shifted so each transit is visible. Shaded symbols denote ZTF-$r$ and LCOGT-$rp$ band data while open symbols denote ZTF-$g$ and LCOGT-$gp$ band data. The vertical grey line at $\mathrm{phase}\,{=}\,0$, with corresponding dates shown, denotes the epoch at which transits are expected to begin based on the 107.2-d period. The observed start of each transit is consistent with the folding period, but transit depths, durations, and shapes appear to vary (see Sec. \ref{subsec:period}). The ZTF data points around 2019 Feb 25 coincide with a predicted transit event, but are too sparsely sampled for conclusive evidence of a transit detection.} \label{fig:photfold}
\end{figure}

In total, we detect two full transit events in the public ZTF data and a third full transit in our LCOGT follow-up observations. A transit egress is also seen at the start of LCOGT observations, around $\mathrm{MJD}\,{=}\,58655$. We constrain the transit spacing by phase-folding all of these data to obtain a close match between the observed transit start times. This occurs with a folding period of $107.2$\,d (see Fig.~\ref{fig:photfold}). Due to the consistent spacing observed between multiple transits, we infer this to be the orbital period of the transiting material.

After phase folding, we note that a few ZTF data points indicating a significant drop in flux lie near to an expected transit event around 2019 Feb 25 (see Fig.~\ref{fig:photfold}). The coverage is very sparse, however, and without a clear indication of transit ingress or egress, we cannot conclude whether these points represent a transit detection. In addition, we searched for transits among the 52 good-quality PS1 data points to further constrain the transit spacing, but find no indication of a transit detection distinguishable from statistical noise ($\Delta \mathrm{m} > 0.15$ mag). Also, when folded at the 107.2\,d period, no PS1 data points overlap with the deepest portions of the ZTF and LCOGT transits.
 
The three full transits observed are variable in both depth and duration. The first transit is the deepest ($\approx40\,\%$) and longest ($\approx25$\,d), while the third transit appears the shallowest ($\approx20\,\%$) and shortest ($\approx15$\,d). The second transit is ${<}\,25$\,d long and sparsely-sampled, but does appear to reach its lowest point much later relative to the first and third transits. Such behavior is reminiscent of WD\,1145+017, whose debris-induced transits exhibit both orbit-to-orbit ($4.5<P_{\mathrm{orb}}<4.9$\,h) and years-long dynamical evolution of their depths, durations, and shapes \citep{Vanderburg2015,Gaensicke2016,Redfield_2017_1,Rappaport2018}. KIC\,8463852, an F-type main sequence star with dust-induced transits, also shows irregularly shaped transits, although it has yet to exhibit conclusively periodic behavior over several years of observations \citep{Boyajian_2016_1,Schaefer_2018_1}. A variety of other objects are known with irregularly shaped transits \citep[see e.g.][]{Rappaport_2019_1}, and are mostly attributed to dusty occulting material.

\section{Discussion} \label{sec:discussion}

\subsection{A Potential Orbital Configuration} \label{sec:configuration}

To account for the presence of transits in ZTF\-J0139+5245, we consider one possible model where a small rocky object was at some point perturbed off its original orbit and brought close enough to the white dwarf to be tidally disrupted into a stream of dust and debris \citep{Debes2002,Jura2003}. Taking the observed ZTF transit spacing of $107.2\,\mathrm{d}$ as the orbital period of that debris, which gives a semi-major axis of $a=76.4\,R_{\odot}\:(0.355\,\mathrm{AU})$, and assuming the material is currently orbiting near to or within the Roche limit \citep[$r_{\mathrm{R}}$,][]{Roche_1849} at periastron, a high eccentricity is required. Adopting for the disrupted body the average density of asteroids \citep[$\rho\,{\approx}\,3.0\,\,\mathrm{g\,cm^{-3}}$,][]{Carry_2012_1}, we find $r_{\mathrm{R}}$ for ZTF\,J0139+5245 to be $1.5\,\mathrm{R}_{\odot}$ \citep{Rappaport2013}. Using $r_{\mathrm{R}}$ as the orbiting material's periastron distance implies an eccentricity of $e>0.97$ and an apastron distance of ${\approx}\,150\,R_{\odot}$ ($0.70\,\mathrm{AU})$.

At periastron, a single dust grain would take ${<}\,$1\,min to transit the white dwarf, and at apastron ${\approx}\,1.5\,$h. Much longer transits are observed by ZTF, suggesting the presence of an extended stream of debris. Large planetesimals may be embedded within this material, but our current constraints on short-timescale transit events from ZTF and McDonald are weak due to sparse coverage.

We estimate a lower limit on the mass of the transiting material that passes between the white dwarf and Earth by first calculating the equivalent width for one of the ZTF transits. The total amount of white dwarf light blocked during the first ZTF transit is equivalent to a total eclipse lasting four days. Since we have no constraints on the density or radial extent of the transiting material, we model this equivalent total eclipse using a flat rectangular cloud composed of non-overlapping, opaque spherical particles. The cloud has a height equal to the white dwarf diameter and a width equal to the equivalent transit duration multiplied by the Keplerian velocity of the orbiting material. If transiting at an orbital distance of 1.5 $R_{\odot}$, the required mass of the occulting material is

\begin{equation}
    M_{d}\approx7\times10^{18} \Big[\frac{r_{d}}{1\:{\mu \mathrm{m}}}\Big] \Big[\frac{\rho_{d}}{2\:\mathrm{g\,cm^{-3}}}\Big] \:\mathrm{g},
\end{equation}

where $r_d$ and $\rho_d$ are the radius and density of a single spherical particle. If transiting at $150\,R_{\odot}$ $(0.70\,\mathrm{AU})$, the mass estimate decreases to

\begin{equation}
    M_{d}\approx7\times10^{16} \Big[\frac{r_{d}}{1\:{\mu \mathrm{m}}}\Big] \Big[\frac{\rho_{d}}{2\:\mathrm{g\,cm^{-3}}}\Big] \:\mathrm{g}.
\end{equation}

In both cases these masses are consistent with asteroid-sized objects, but represent only the observed transiting material. Also, relaxing the various assumptions we made would tend to increase the estimated mass. For instance, our assumption of non-overlapping particles works well in the optically-thin limit, but for the first ZTF transit we estimate a roughly 15\,\% increase in mass when allowing for overlapping particles. We consider this a small effect, however, compared to the two orders-of-magnitude difference seen when choosing different orbital distances during transit. Our estimate also neglects mass contributed by gas in the system.

\subsection{Possible Origins of the Transiting Material} \label{sec:origins}

The origin of the transiting material in ZTF\-J0139+5245 faces many of the same questions which have been posed for metal-polluted white dwarfs in general. During the post-main sequence lifetime of the progenitor star, any planetary objects inside of at least 1.5\,AU are likely to be engulfed and destroyed \citep{Mustill_2012_1}, and only objects with as much mass as a brown dwarf are expected to survive engulfment \citep{Livio_1984_1,Soker_1984_1,Nelemans_1998_1,Maxted_2006_1}. To account for the prevalence of metal polluted white dwarfs, a mechanism is needed which can bring rocky planetary material to the surface of the white dwarf from outside the engulfment region. The canonical tidal disruption model provides such a mechanism, whereby asteroids are gravitationally perturbed to highly eccentric, star-grazing orbits following the orbital expansion of outer planets due to stellar mass loss \citep{Debes2002,Jura2003}. Subsequent works have shown that secular perturbations on asteroids work on timescales that match the observed age distribution of metal polluted white dwarfs \citep{Frewen_2014_1,Mustill_2018_1,Smallwood_2018_1}.

The orbital period of the transiting material in ZTF\,J0139+5245 clearly places it within the region previously engulfed during post-main sequence evolution, so again a mechanism is required which places material at this location. With irregularly shaped transits and Ca\,{\sc ii} K and H absorption lines of photospheric or circumstellar origin, the current observations are consistent with the standard tidal disruption model. However, given the long orbital period of the transiting material, this model requires a very high eccentricity to bring the material near to or within the Roche limit (see Sec.~\ref{sec:configuration}). 

Such high eccentricities are expected during the initial infall and disruption event \citep{Debes_2012_1,Veras2014}, but confirming a highly eccentric orbit is not possible with the observations presented here. Thus, the standard tidal disruption model remains just one possibility for the origin of the observed transiting material. If confirmed, however, this model would provide a natural explanation for the potential variability observed in Ca\,{\sc ii} line strengths. Each pass at periastron may result in a gas production event due to sublimation or collisions so that the transiting material remains supplied with metallic gas. This way, an increased column density of metallic gas can be observed during transit, potentially accompanied by increased photospheric accretion.

Another possibility is that the transiting material is somehow related to an object that survived post-main sequence engulfment. While objects less massive than brown dwarfs are typically expected to be destroyed, the recent detection of a giant planet in close orbit around WD\,J091405.30+191412.25 has shown that massive planets down to roughly $1\,M_J$ may survive engulfment as well \citep{Gaensicke_2019_1}.  Using the measured \textit{J}-band magnitude from UKIRT \citep{Dye_2019_1}, we can place limits on the presence of a companion. We find that $J\,{=}\,19.10\pm0.07$ (AB mag) is consistent with the synthetic \textit{J}-band magnitude \citep{Holberg_2016_1} of an isolated white dwarf (see Fig.~\ref{fig:spectrum}). Using the absolute \textit{J}-band magnitudes for L-dwarfs from \citet{Dupuy_2012_1}, we can exclude the presence of any companion with a spectral type earlier than L6.  

The orbital periods of post-common-envelope binaries with such low-mass companions are typically in the range of 0.1--1.0\,d \citep{Nebot_2011_1,Zorotovic_2014_1,Camacho_2014_1} and, as shown in Sec.~\ref{sec:configuration}, the observed transit durations cannot be accounted for by a single solid object. In addition, the vast majority of metal-polluted white dwarfs with a detected infrared excess are single stars \citep{Wilson_2019_1}.  With these constraints, we find a brown dwarf or planetary mass companion that has survived common envelope evolution to be an unlikely source for the observed transits.

Radial velocity measurements throughout the orbit would place additional constraints on any companion mass, though even at the upper end of brown dwarf masses ($0.08\,M_{\odot}$), radial velocity semi-amplitudes of just ${\sim}\,5\,\mathrm{km\,s^{-1}}$ would be expected for an edge-on, circular orbit at the observed orbital period. The uncertainties in our current spectroscopic observations are too large to further rule out the presence of a companion.

Another mechanism which may transport planetary objects into the previously cleared-out region is the late unpacking of densely-spaced planetary systems \citep{Veras_2015_2}. Planet incursions alone cannot account for the 15--25\,d transit durations observed, so a planet-planet or planet-asteroid collision must also occur to create an extended stream of planetary debris. The likelihoods of such collisions are poorly constrained, though \citet{Veras_2015_2} state that $25\,\%$ of terrestrial planet instabilities in their simulations are planet-planet collisions, and could be a plausible way to account for some heavily metal-polluted white dwarfs.

Lastly, regardless of how the observed planetary material was brought onto its current orbit around ZTF\,J0139+5245, there exist mechanisms beyond tidal disruption and collisions which may produce an extended stream of planetary debris. One possibility is the rotational fission of an aspherical asteroid \citep{Makarov2019,Veras_2020_1}. This mechanism does not require the asteroid to orbit within the Roche limit, though \citet{Veras_2020_1} found that an asteroid around ZTF\,J0139+5245 must still achieve pericenter distances of 1--3$\,r_{\mathrm{R}}$ for rotational breakup to occur, depending on how elongated the asteroid is.

\section{Conclusions} \label{sec:conclusions}

We have presented the discovery of a white dwarf exhibiting photometric evidence for transits caused by planetary material orbiting far outside the Roche limit. We have placed loose constraints on the transit recurrence time (${\approx}\,107.2$\,d) and found transit-to-transit variations in depth (20--45\,\%), duration (15--25\,d), and shape. We obtained the first optical spectra of ZTF\,J0139+5245 and identified the white dwarf as a DA due to strong Balmer lines, with $T_{\mathrm{eff}}^{3\mathrm{D}}=10{,}530\pm140\,\mathrm{K}$ and $\log(g)^{3\mathrm{D}}=7.86\pm0.06$. 

We detected Ca\,{\sc ii}~H \& K absorption features that are likely the result of photospheric metal accretion or an increase in circumstellar metallic gas along the line of sight. The strongest Ca lines were observed at 15--20\% transit depth, suggesting a correlation between transit events and increased calcium absorption. The change in Ca\,{\sc ii}~K line strength is only 2-sigma significant in our data, however, so the variability of Ca absorption and its correlation with transit events needs verification with high-resolution spectroscopy over the entire orbital period. Additionally, {\em James Webb Space Telescope} ({\em JWST}) observations are needed for infrared flux constraints. Detection of an infrared excess could further rule out the possibility of a sub-stellar companion, help measure the temperature of the circumstellar dust disk, and potentially reveal long-term infrared variability of the system.

While the observations presented here are consistent with the canonical tidal disruption model, the long orbital period requires a very high eccentricity ($e\,{>}\,0.97$) which cannot yet be confirmed. Other mechanisms such as the rotational fission of aspherical asteroids \citep{Makarov2019,Veras_2020_1}, late planetary system unpacking \citep{Veras_2015_2}, and collisions remain possible sources of the transiting debris, but we find an object which has survived post-main sequence engulfment to be an unlikely source. If tidal disruption is indeed responsible for the observed transiting debris in ZTF\,J0139+5245, the long orbital period observed may suggest a much earlier phase of tidal disruption compared to WD\,1145+017, whose transiting debris is also consistent with the tidal disruption model. Long term monitoring of the transit periodicity may provide useful insight into the physical processes governing the evolution of its orbiting debris, such as Poynting-Robertson drag, collisions, and sublimation  \citep{Veras_2015_1,Farihi_2016_1,Kenyon_2017_1}. Lastly, we expect a more systematic search of a larger space volume of white dwarfs in ZTF, and in future surveys such as the Large Synoptic Survey Telescope, will yield many more discoveries like ZTF\,J0139+5245.

\acknowledgments
The authors would like to thank J. Farihi for providing useful feedback on a draft of this Letter, D. Koester for the use of his models, and T. Marsh for developing and making available both {\sc pamela} and {\sc molly}. Data from McDonald Observatory were obtained with financial support from NASA K2 Cycle 5 Grant 80NSSC18K0387 and the Wooten Center for Astrophysical Plasma Properties (WCAPP) under DOE grant DE-FOA-0001634.
PI acknowledges financial support from the Spanish Ministry of Economy and Competitiveness (MINECO) under the 2015 Severo Ochoa Programme MINECO SEV–2015–0548.
BTG was supported by the UK STFC grant ST/P000495.
This material is based upon work supported by the National Science Foundation under Award AST-1903828.
The research leading to these results has received funding
from the European Research Council under the European
Union’s Horizon 2020 research and innovation programme
n. 677706 (WD3D).

This work is based on observations obtained with the Samuel Oschin 48-inch Telescope at the Palomar Observatory as part of the Zwicky Transient Facility project which is supported by NSF Grant No.\ AST-1440341 and the participating institutions of the ZTF collaboration. 
We make use of observations and reduction software from the LCOGT network,
and include data from the Pan-STARRS1 Surveys (PS1) and the PS1 public science archive which have been made possible through contributions by participating institutions of the PS1 collaboration (\url{https://panstarrs.stsci.edu/}).
This work makes use of the synthetic white dwarf colors provided by Pierre Bergeron (\url{http://www.astro.umontreal.ca/~bergeron/CoolingModels})

We also make use of data from the United Kingdom Infra-Red Telescope from the year 2016 (UKIRT, \url{http://www.ukirt.hawaii.edu/}), the NASA Galaxy Evolution Explorer ({\em GALEX}, \url{http://www.galex.caltech.edu/}), the SDSS-III (\url{http://www.sdss3.org/}), and from the European Space Agency (ESA) mission {\em Gaia} (\url{https://www.cosmos.esa.int/gaia}) processed by the {\em Gaia} Data Processing and Analysis Consortium (DPAC, \url{https://www.cosmos.esa.int/web/gaia/dpac/consortium}).

\textit{Additional Software/Resources:} Astropy \citep{Astropy_2013, Astropy_2018}, Photoutils \citep{Bradley_2019_2533376}, {\sc iraf} (distributed by the National Optical Astronomy Observatory, which is operated by the Association of Universities for Research in Astronomy (AURA) under a cooperative agreement with the National Science Foundation.), the NASA Astrophysics Data System (ADS), and SIMBAD and VizieR (operated at CDS, Strasbourg, France).

\bibliography{ref}

\begin{thebibliography}{}
\expandafter\ifx\csname natexlab\endcsname\relax\def\natexlab#1{#1}\fi
\providecommand{\url}[1]{\href{#1}{#1}}

\bibitem[{{Ahn} {et~al.}(2012){Ahn}, {Alexandroff}, {Allende Prieto},
  {Anderson}, {Anderton}, {Andrews}, {Aubourg}, {Bailey}, {Balbinot}, {Barnes},
  \& et~al.}]{Ahn2012}
{Ahn}, C.~P., {Alexandroff}, R., {Allende Prieto}, C., {et~al.} 2012, \apjs,
  203, 21

\bibitem[{{Astropy Collaboration} {et~al.}(2013){Astropy Collaboration},
  {Robitaille}, {Tollerud}, {Greenfield}, {Droettboom}, {Bray}, {Aldcroft},
  {Davis}, {Ginsburg}, {Price-Whelan}, {Kerzendorf}, {Conley}, {Crighton},
  {Barbary}, {Muna}, {Ferguson}, {Grollier}, {Parikh}, {Nair}, {Unther},
  {Deil}, {Woillez}, {Conseil}, {Kramer}, {Turner}, {Singer}, {Fox}, {Weaver},
  {Zabalza}, {Edwards}, {Azalee Bostroem}, {Burke}, {Casey}, {Crawford},
  {Dencheva}, {Ely}, {Jenness}, {Labrie}, {Lim}, {Pierfederici}, {Pontzen},
  {Ptak}, {Refsdal}, {Servillat}, \& {Streicher}}]{Astropy_2013}
{Astropy Collaboration}, {Robitaille}, T.~P., {Tollerud}, E.~J., {et~al.} 2013,
  \aap, 558, A33

\bibitem[{{Astropy Collaboration} {et~al.}(2018){Astropy Collaboration},
  {Price-Whelan}, {Sip{\H{o}}cz}, {G{\"u}nther}, {Lim}, {Crawford}, {Conseil},
  {Shupe}, {Craig}, {Dencheva}, {Ginsburg}, {Vand erPlas}, {Bradley},
  {P{\'e}rez-Su{\'a}rez}, {de Val-Borro}, {Aldcroft}, {Cruz}, {Robitaille},
  {Tollerud}, {Ardelean}, {Babej}, {Bach}, {Bachetti}, {Bakanov}, {Bamford},
  {Barentsen}, {Barmby}, {Baumbach}, {Berry}, {Biscani}, {Boquien}, {Bostroem},
  {Bouma}, {Brammer}, {Bray}, {Breytenbach}, {Buddelmeijer}, {Burke},
  {Calderone}, {Cano Rodr{\'\i}guez}, {Cara}, {Cardoso}, {Cheedella}, {Copin},
  {Corrales}, {Crichton}, {D'Avella}, {Deil}, {Depagne}, {Dietrich}, {Donath},
  {Droettboom}, {Earl}, {Erben}, {Fabbro}, {Ferreira}, {Finethy}, {Fox},
  {Garrison}, {Gibbons}, {Goldstein}, {Gommers}, {Greco}, {Greenfield},
  {Groener}, {Grollier}, {Hagen}, {Hirst}, {Homeier}, {Horton}, {Hosseinzadeh},
  {Hu}, {Hunkeler}, {Ivezi{\'c}}, {Jain}, {Jenness}, {Kanarek}, {Kendrew},
  {Kern}, {Kerzendorf}, {Khvalko}, {King}, {Kirkby}, {Kulkarni}, {Kumar},
  {Lee}, {Lenz}, {Littlefair}, {Ma}, {Macleod}, {Mastropietro}, {McCully},
  {Montagnac}, {Morris}, {Mueller}, {Mumford}, {Muna}, {Murphy}, {Nelson},
  {Nguyen}, {Ninan}, {N{\"o}the}, {Ogaz}, {Oh}, {Parejko}, {Parley}, {Pascual},
  {Patil}, {Patil}, {Plunkett}, {Prochaska}, {Rastogi}, {Reddy Janga},
  {Sabater}, {Sakurikar}, {Seifert}, {Sherbert}, {Sherwood-Taylor}, {Shih},
  {Sick}, {Silbiger}, {Singanamalla}, {Singer}, {Sladen}, {Sooley},
  {Sornarajah}, {Streicher}, {Teuben}, {Thomas}, {Tremblay}, {Turner},
  {Terr{\'o}n}, {van Kerkwijk}, {de la Vega}, {Watkins}, {Weaver}, {Whitmore},
  {Woillez}, {Zabalza}, \& {Astropy Contributors}}]{Astropy_2018}
{Astropy Collaboration}, {Price-Whelan}, A.~M., {Sip{\H{o}}cz}, B.~M., {et~al.}
  2018, \aj, 156, 123

\bibitem[{{Bailer-Jones} {et~al.}(2018){Bailer-Jones}, {Rybizki}, {Fouesneau},
  {Mantelet}, \& {Andrae}}]{Bailer-Jones2018}
{Bailer-Jones}, C.~A.~L., {Rybizki}, J., {Fouesneau}, M., {Mantelet}, G., \&
  {Andrae}, R. 2018, \aj, 156, 58

\bibitem[{{Bellm} {et~al.}(2019){Bellm}, {Kulkarni}, {Graham}, {Dekany},
  {Smith}, {Riddle}, {Masci}, {Helou}, {Prince}, {Adams}, {Barbarino},
  {Barlow}, {Bauer}, {Beck}, {Belicki}, {Biswas}, {Blagorodnova}, {Bodewits},
  {Bolin}, {Brinnel}, {Brooke}, {Bue}, {Bulla}, {Burruss}, {Cenko}, {Chang},
  {Connolly}, {Coughlin}, {Cromer}, {Cunningham}, {De}, {Delacroix}, {Desai},
  {Duev}, {Eadie}, {Farnham}, {Feeney}, {Feindt}, {Flynn}, {Franckowiak},
  {Frederick}, {Fremling}, {Gal-Yam}, {Gezari}, {Giomi}, {Goldstein},
  {Golkhou}, {Goobar}, {Groom}, {Hacopians}, {Hale}, {Henning}, {Ho}, {Hover},
  {Howell}, {Hung}, {Huppenkothen}, {Imel}, {Ip}, {Ivezi{\'c}}, {Jackson},
  {Jones}, {Juric}, {Kasliwal}, {Kaspi}, {Kaye}, {Kelley}, {Kowalski},
  {Kramer}, {Kupfer}, {Landry}, {Laher}, {Lee}, {Lin}, {Lin}, {Lunnan},
  {Giomi}, {Mahabal}, {Mao}, {Miller}, {Monkewitz}, {Murphy}, {Ngeow},
  {Nordin}, {Nugent}, {Ofek}, {Patterson}, {Penprase}, {Porter}, {Rauch},
  {Rebbapragada}, {Reiley}, {Rigault}, {Rodriguez}, {van Roestel}, {Rusholme},
  {van Santen}, {Schulze}, {Shupe}, {Singer}, {Soumagnac}, {Stein}, {Surace},
  {Sollerman}, {Szkody}, {Taddia}, {Terek}, {Van Sistine}, {van Velzen},
  {Vestrand}, {Walters}, {Ward}, {Ye}, {Yu}, {Yan}, \& {Zolkower}}]{Bellm2019}
{Bellm}, E.~C., {Kulkarni}, S.~R., {Graham}, M.~J., {et~al.} 2019, \pasp, 131,
  018002

\bibitem[{{Boyajian} {et~al.}(2016){Boyajian}, {LaCourse}, {Rappaport},
  {Fabrycky}, {Fischer}, {Gandolfi}, {Kennedy}, {Korhonen}, {Liu}, {Moor},
  {Olah}, {Vida}, {Wyatt}, {Best}, {Brewer}, {Ciesla}, {Cs{\'a}k}, {Deeg},
  {Dupuy}, {Handler}, {Heng}, {Howell}, {Ishikawa}, {Kov{\'a}cs}, {Kozakis},
  {Kriskovics}, {Lehtinen}, {Lintott}, {Lynn}, {Nespral}, {Nikbakhsh},
  {Schawinski}, {Schmitt}, {Smith}, {Szabo}, {Szabo}, {Viuho}, {Wang},
  {Weiksnar}, {Bosch}, {Connors}, {Goodman}, {Green}, {Hoekstra}, {Jebson},
  {Jek}, {Omohundro}, {Schwengeler}, \& {Szewczyk}}]{Boyajian_2016_1}
{Boyajian}, T.~S., {LaCourse}, D.~M., {Rappaport}, S.~A., {et~al.} 2016,
  \mnras, 457, 3988

\bibitem[{Bradley {et~al.}(2019)Bradley, Sip{\H o}cz, Robitaille, Tollerud,
  Vin{\'{\i}}cius, Deil, Barbary, G{\"u}nther, Cara, Busko, Conseil,
  Droettboom, Bostroem, Bray, Bratholm, Wilson, Craig, Barentsen, Pascual,
  Donath, Greco, Perren, Lim, \& Kerzendorf}]{Bradley_2019_2533376}
Bradley, L., Sip{\H o}cz, B., Robitaille, T., {et~al.} 2019, astropy/photutils:
  v0.6, , , doi:10.5281/zenodo.2533376.
\newblock \url{https://doi.org/10.5281/zenodo.2533376}

\bibitem[{{Breger} {et~al.}(1993){Breger}, {Stich}, {Garrido}, {Martin},
  {Jiang}, {Li}, {Hube}, {Ostermann}, {Paparo}, \& {Scheck}}]{Breger_1993_1}
{Breger}, M., {Stich}, J., {Garrido}, R., {et~al.} 1993, \aap, 271, 482

\bibitem[{{Brown} {et~al.}(2013){Brown}, {Baliber}, {Bianco}, {Bowman},
  {Burleson}, {Conway}, {Crellin}, {Depagne}, {De Vera}, {Dilday}, {Dragomir},
  {Dubberley}, {Eastman}, {Elphick}, {Falarski}, {Foale}, {Ford}, {Fulton},
  {Garza}, {Gomez}, {Graham}, {Greene}, {Haldeman}, {Hawkins}, {Haworth},
  {Haynes}, {Hidas}, {Hjelstrom}, {Howell}, {Hygelund}, {Lister}, {Lobdill},
  {Martinez}, {Mullins}, {Norbury}, {Parrent}, {Paulson}, {Petry}, {Pickles},
  {Posner}, {Rosing}, {Ross}, {Sand}, {Saunders}, {Shobbrook}, {Shporer},
  {Street}, {Thomas}, {Tsapras}, {Tufts}, {Valenti}, {Vander Horst}, {Walker},
  {White}, \& {Willis}}]{Brown_2013_1}
{Brown}, T.~M., {Baliber}, N., {Bianco}, F.~B., {et~al.} 2013, \pasp, 125, 1031

\bibitem[{{Camacho} {et~al.}(2014){Camacho}, {Torres}, {Garc{\'\i}a-Berro},
  {Zorotovic}, {Schreiber}, {Rebassa-Mansergas}, {Nebot G{\'o}mez-Mor{\'a}n},
  \& {G{\"a}nsicke}}]{Camacho_2014_1}
{Camacho}, J., {Torres}, S., {Garc{\'\i}a-Berro}, E., {et~al.} 2014, \aap, 566,
  A86

\bibitem[{{Capitanio} {et~al.}(2017){Capitanio}, {Lallement}, {Vergely},
  {Elyajouri}, \& {Monreal-Ibero}}]{Capitanio_2017_1}
{Capitanio}, L., {Lallement}, R., {Vergely}, J.~L., {Elyajouri}, M., \&
  {Monreal-Ibero}, A. 2017, \aap, 606, A65

\bibitem[{{Carry}(2012)}]{Carry_2012_1}
{Carry}, B. 2012, \planss, 73, 98

\bibitem[{{Chambers} {et~al.}(2016){Chambers}, {Magnier}, {Metcalfe},
  {Flewelling}, {Huber}, {Waters}, {Denneau}, {Draper}, {Farrow}, \&
  {Finkbeiner}}]{Chambers2016}
{Chambers}, K.~C., {Magnier}, E.~A., {Metcalfe}, N., {et~al.} 2016, arXiv
  e-prints, arXiv:1612.05560

\bibitem[{{Cunningham} {et~al.}(2019){Cunningham}, {Tremblay}, {Freytag},
  {Ludwig}, \& {Koester}}]{Cunningham_2019_1}
{Cunningham}, T., {Tremblay}, P.-E., {Freytag}, B., {Ludwig}, H.-G., \&
  {Koester}, D. 2019, \mnras, 488, 2503

\bibitem[{{Debes} \& {L{\'o}pez-Morales}(2008)}]{Debes_2008_1}
{Debes}, J.~H., \& {L{\'o}pez-Morales}, M. 2008, \apjl, 677, L43

\bibitem[{{Debes} \& {Sigurdsson}(2002)}]{Debes2002}
{Debes}, J.~H., \& {Sigurdsson}, S. 2002, \apj, 572, 556

\bibitem[{{Debes} {et~al.}(2012){Debes}, {Walsh}, \& {Stark}}]{Debes_2012_1}
{Debes}, J.~H., {Walsh}, K.~J., \& {Stark}, C. 2012, \apj, 747, 148

\bibitem[{{Dupuy} \& {Liu}(2012)}]{Dupuy_2012_1}
{Dupuy}, T.~J., \& {Liu}, M.~C. 2012, \apjs, 201, 19

\bibitem[{{Dye} {et~al.}(2018){Dye}, {Lawrence}, {Read}, {Fan}, {Kerr},
  {Varricatt}, {Furnell}, {Edge}, {Irwin}, {Hambly}, {Lucas}, {Almaini},
  {Chambers}, {Green}, {Hewett}, {Liu}, {McGreer}, {Best}, {Zhang}, {Sutorius},
  {Froebrich}, {Magnier}, {Hasinger}, {Lederer}, {Bold}, \&
  {Tedds}}]{Dye_2019_1}
{Dye}, S., {Lawrence}, A., {Read}, M.~A., {et~al.} 2018, \mnras, 473, 5113

\bibitem[{{Evans} {et~al.}(2018){Evans}, {Riello}, {De Angeli}, {Carrasco},
  {Montegriffo}, {Fabricius}, {Jordi}, {Palaversa}, {Diener}, \&
  {Busso}}]{Evans2018}
{Evans}, D.~W., {Riello}, M., {De Angeli}, F., {et~al.} 2018, \aap, 616, A4

\bibitem[{{Farihi}(2016)}]{Farihi_2016_1}
{Farihi}, J. 2016, \nar, 71, 9

\bibitem[{{Fitzpatrick}(1999)}]{Fitzpatrick1999}
{Fitzpatrick}, E.~L. 1999, \pasp, 111, 63

\bibitem[{{Fontaine} {et~al.}(2001){Fontaine}, {Brassard}, \&
  {Bergeron}}]{Fontaine2001}
{Fontaine}, G., {Brassard}, P., \& {Bergeron}, P. 2001, \pasp, 113, 409

\bibitem[{{Frewen} \& {Hansen}(2014)}]{Frewen_2014_1}
{Frewen}, S.~F.~N., \& {Hansen}, B.~M.~S. 2014, \mnras, 439, 2442

\bibitem[{{Fulton} {et~al.}(2014){Fulton}, {Tonry}, {Flewelling}, {Burgett},
  {Chambers}, {Hodapp}, {Huber}, {Kaiser}, {Wainscoat}, \&
  {Waters}}]{Fulton2014}
{Fulton}, B.~J., {Tonry}, J.~L., {Flewelling}, H., {et~al.} 2014, \apj, 796,
  114

\bibitem[{{Gaia Collaboration} {et~al.}(2018){Gaia Collaboration}, {Brown},
  {Vallenari}, {Prusti}, {de Bruijne}, {Babusiaux}, {Bailer-Jones}, {Biermann},
  {Evans}, \& {Eyer}}]{Gaia2}
{Gaia Collaboration}, {Brown}, A.~G.~A., {Vallenari}, A., {et~al.} 2018, \aap,
  616, A1

\bibitem[{{G{\"a}nsicke} {et~al.}(2019){G{\"a}nsicke}, {Schreiber}, {Toloza},
  {Fusillo}, {Koester}, \& {Manser}}]{Gaensicke_2019_1}
{G{\"a}nsicke}, B.~T., {Schreiber}, M.~R., {Toloza}, O., {et~al.} 2019, \nat,
  576, 61

\bibitem[{{G{\"a}nsicke} {et~al.}(2016){G{\"a}nsicke}, {Aungwerojwit}, {Marsh},
  {Dhillon}, {Sahman}, {Veras}, {Farihi}, {Chote}, {Ashley}, {Arjyotha},
  {Rattanasoon}, {Littlefair}, {Pollacco}, \& {Burleigh}}]{Gaensicke2016}
{G{\"a}nsicke}, B.~T., {Aungwerojwit}, A., {Marsh}, T.~R., {et~al.} 2016,
  \apjl, 818, L7

\bibitem[{{Gentile Fusillo} {et~al.}(2019){Gentile Fusillo}, {Tremblay},
  {G{\"a}nsicke}, {Manser}, {Cunningham}, {Cukanovaite}, {Hollands}, {Marsh},
  {Raddi}, \& {Jordan}}]{GF2019}
{Gentile Fusillo}, N.~P., {Tremblay}, P.-E., {G{\"a}nsicke}, B.~T., {et~al.}
  2019, \mnras, 482, 4570

\bibitem[{{Gianninas} {et~al.}(2015){Gianninas}, {Kilic}, {Brown}, {Canton}, \&
  {Kenyon}}]{Gianninas2015}
{Gianninas}, A., {Kilic}, M., {Brown}, W.~R., {Canton}, P., \& {Kenyon}, S.~J.
  2015, \apj, 812, 167

\bibitem[{{Girven} {et~al.}(2011){Girven}, {G{\"a}nsicke}, {Steeghs}, \&
  {Koester}}]{Girven_2011_1}
{Girven}, J., {G{\"a}nsicke}, B.~T., {Steeghs}, D., \& {Koester}, D. 2011,
  \mnras, 417, 1210

\bibitem[{{Hewett} {et~al.}(2006){Hewett}, {Warren}, {Leggett}, \&
  {Hodgkin}}]{Hewett_2006_1}
{Hewett}, P.~C., {Warren}, S.~J., {Leggett}, S.~K., \& {Hodgkin}, S.~T. 2006,
  \mnras, 367, 454

\bibitem[{{Holberg} \& {Bergeron}(2006)}]{Holberg_2016_1}
{Holberg}, J.~B., \& {Bergeron}, P. 2006, \aj, 132, 1221

\bibitem[{{Horne}(1986)}]{Horne1986-1}
{Horne}, K. 1986, \pasp, 98, 609

\bibitem[{{Indebetouw} {et~al.}(2005){Indebetouw}, {Mathis}, {Babler}, {Meade},
  {Watson}, {Whitney}, {Wolff}, {Wolfire}, {Cohen}, {Bania}, {Benjamin},
  {Clemens}, {Dickey}, {Jackson}, {Kobulnicky}, {Marston}, {Mercer},
  {Stauffer}, {Stolovy}, \& {Churchwell}}]{Indebetouw2005}
{Indebetouw}, R., {Mathis}, J.~S., {Babler}, B.~L., {et~al.} 2005, \apj, 619,
  931

\bibitem[{{Jura}(2003)}]{Jura2003}
{Jura}, M. 2003, \apj, 584, L91

\bibitem[{{Kaiser} {et~al.}(2010){Kaiser}, {Burgett}, {Chambers}, {Denneau},
  {Heasley}, {Jedicke}, {Magnier}, {Morgan}, {Onaka}, \& {Tonry}}]{Kaiser2010}
{Kaiser}, N., {Burgett}, W., {Chambers}, K., {et~al.} 2010, in Society of
  Photo-Optical Instrumentation Engineers (SPIE) Conference Series, Vol. 7733,
  \procspie, 77330E

\bibitem[{{Kanaan} {et~al.}(2002){Kanaan}, {Kepler}, \& {Winget}}]{Kanaan2002}
{Kanaan}, A., {Kepler}, S.~O., \& {Winget}, D.~E. 2002, \aap, 389, 896

\bibitem[{{Kenyon} \& {Bromley}(2017)}]{Kenyon_2017_1}
{Kenyon}, S.~J., \& {Bromley}, B.~C. 2017, \apj, 844, 116

\bibitem[{{Koester}(2010)}]{Koester_2010_1}
{Koester}, D. 2010, \memsai, 81, 921

\bibitem[{{Koester} {et~al.}(2014){Koester}, {G{\"a}nsicke}, \&
  {Farihi}}]{Koester2014}
{Koester}, D., {G{\"a}nsicke}, B.~T., \& {Farihi}, J. 2014, \aap, 566, A34

\bibitem[{{Koester} \& {Wilken}(2006)}]{Koester_2006_1}
{Koester}, D., \& {Wilken}, D. 2006, \aap, 453, 1051

\bibitem[{{Kuschnig} {et~al.}(1997){Kuschnig}, {Weiss}, {Gruber}, {Bely}, \&
  {Jenkner}}]{Kuschnig_1997_1}
{Kuschnig}, R., {Weiss}, W.~W., {Gruber}, R., {Bely}, P.~Y., \& {Jenkner}, H.
  1997, \aap, 328, 544

\bibitem[{{Liebert} {et~al.}(2005){Liebert}, {Bergeron}, \&
  {Holberg}}]{Liebert2005}
{Liebert}, J., {Bergeron}, P., \& {Holberg}, J.~B. 2005, \apjs, 156, 47

\bibitem[{{Lindegren} {et~al.}(2018){Lindegren}, {Hern{\'a}ndez}, {Bombrun},
  {Klioner}, {Bastian}, {Ramos-Lerate}, {de Torres}, {Steidelm{\"u}ller},
  {Stephenson}, \& {Hobbs}}]{Lindegren2018}
{Lindegren}, L., {Hern{\'a}ndez}, J., {Bombrun}, A., {et~al.} 2018, \aap, 616,
  A2

\bibitem[{{Livio} \& {Soker}(1984)}]{Livio_1984_1}
{Livio}, M., \& {Soker}, N. 1984, \mnras, 208, 763

\bibitem[{{Magnier} {et~al.}(2013){Magnier}, {Schlafly}, {Finkbeiner}, {Juric},
  {Tonry}, {Burgett}, {Chambers}, {Flewelling}, {Kaiser}, {Kudritzki},
  {Morgan}, {Price}, {Sweeney}, \& {Stubbs}}]{Magnier2013}
{Magnier}, E.~A., {Schlafly}, E., {Finkbeiner}, D., {et~al.} 2013, \apjs, 205,
  20

\bibitem[{{Makarov} \& {Veras}(2019)}]{Makarov2019}
{Makarov}, V.~V., \& {Veras}, D. 2019, arXiv e-prints, arXiv:1908.04612

\bibitem[{{Marsh}(1989)}]{Marsh1989}
{Marsh}, T.~R. 1989, \pasp, 101, 1032

\bibitem[{{Martin} {et~al.}(2005){Martin}, {Fanson}, {Schiminovich},
  {Morrissey}, {Friedman}, {Barlow}, {Conrow}, {Grange}, {Jelinsky},
  {Milliard}, {Siegmund}, {Bianchi}, {Byun}, {Donas}, {Forster}, {Heckman},
  {Lee}, {Madore}, {Malina}, {Neff}, {Rich}, {Small}, {Surber}, {Szalay},
  {Welsh}, \& {Wyder}}]{Martin2005}
{Martin}, D.~C., {Fanson}, J., {Schiminovich}, D., {et~al.} 2005, \apjl, 619,
  L1

\bibitem[{{Masci} {et~al.}(2019){Masci}, {Laher}, {Rusholme}, {Shupe}, {Groom},
  {Surace}, {Jackson}, {Monkewitz}, {Beck}, \& {Flynn}}]{Masci2019}
{Masci}, F.~J., {Laher}, R.~R., {Rusholme}, B., {et~al.} 2019, \pasp, 131,
  018003

\bibitem[{{Maxted} {et~al.}(2006){Maxted}, {Napiwotzki}, {Dobbie}, \&
  {Burleigh}}]{Maxted_2006_1}
{Maxted}, P.~F.~L., {Napiwotzki}, R., {Dobbie}, P.~D., \& {Burleigh}, M.~R.
  2006, \nat, 442, 543

\bibitem[{{Morrissey} {et~al.}(2005){Morrissey}, {Schiminovich}, {Barlow},
  {Martin}, {Blakkolb}, {Conrow}, {Cooke}, {Erickson}, {Fanson}, {Friedman},
  {Grange}, {Jelinsky}, {Lee}, {Liu}, {Mazer}, {McLean}, {Milliard}, {Randall},
  {Schmitigal}, {Sen}, {Siegmund}, {Surber}, {Vaughan}, {Viton}, {Welsh},
  {Bianchi}, {Byun}, {Donas}, {Forster}, {Heckman}, {Lee}, {Madore}, {Malina},
  {Neff}, {Rich}, {Small}, {Szalay}, \& {Wyder}}]{Morrissey2005}
{Morrissey}, P., {Schiminovich}, D., {Barlow}, T.~A., {et~al.} 2005, \apjl,
  619, L7

\bibitem[{{Mukadam} {et~al.}(2006){Mukadam}, {Montgomery}, {Winget}, {Kepler},
  \& {Clemens}}]{Mukadam_2006_1}
{Mukadam}, A.~S., {Montgomery}, M.~H., {Winget}, D.~E., {Kepler}, S.~O., \&
  {Clemens}, J.~C. 2006, \apj, 640, 956

\bibitem[{{Mustill} \& {Villaver}(2012)}]{Mustill_2012_1}
{Mustill}, A.~J., \& {Villaver}, E. 2012, \apj, 761, 121

\bibitem[{{Mustill} {et~al.}(2018){Mustill}, {Villaver}, {Veras},
  {G{\"a}nsicke}, \& {Bonsor}}]{Mustill_2018_1}
{Mustill}, A.~J., {Villaver}, E., {Veras}, D., {G{\"a}nsicke}, B.~T., \&
  {Bonsor}, A. 2018, \mnras, 476, 3939

\bibitem[{{Nebot G{\'o}mez-Mor{\'a}n} {et~al.}(2011){Nebot
  G{\'o}mez-Mor{\'a}n}, {G{\"a}nsicke}, {Schreiber}, {Rebassa-Mansergas},
  {Schwope}, {Southworth}, {Aungwerojwit}, {Bothe}, {Davis}, {Kolb},
  {M{\"u}ller}, {Papadaki}, {Pyrzas}, {Rabitz}, {Rodr{\'\i}guez-Gil},
  {Schmidtobreick}, {Schwarz}, {Tappert}, {Toloza}, {Vogel}, \&
  {Zorotovic}}]{Nebot_2011_1}
{Nebot G{\'o}mez-Mor{\'a}n}, A., {G{\"a}nsicke}, B.~T., {Schreiber}, M.~R.,
  {et~al.} 2011, \aap, 536, A43

\bibitem[{{Nelemans} \& {Tauris}(1998)}]{Nelemans_1998_1}
{Nelemans}, G., \& {Tauris}, T.~M. 1998, \aap, 335, L85

\bibitem[{{Patterson} {et~al.}(2019){Patterson}, {Bellm}, {Rusholme}, {Masci},
  {Juric}, {Krughoff}, {Golkhou}, {Graham}, {Kulkarni}, {Helou}, \& {Zwicky
  Transient Facility Collaboration}}]{Patterson2019}
{Patterson}, M.~T., {Bellm}, E.~C., {Rusholme}, B., {et~al.} 2019, \pasp, 131,
  018001

\bibitem[{{Rappaport} {et~al.}(2018){Rappaport}, {Gary}, {Vanderburg}, {Xu},
  {Pooley}, \& {Mukai}}]{Rappaport2018}
{Rappaport}, S., {Gary}, B.~L., {Vanderburg}, A., {et~al.} 2018, \mnras, 474,
  933

\bibitem[{{Rappaport} {et~al.}(2013){Rappaport}, {Sanchis-Ojeda}, {Rogers},
  {Levine}, \& {Winn}}]{Rappaport2013}
{Rappaport}, S., {Sanchis-Ojeda}, R., {Rogers}, L.~A., {Levine}, A., \& {Winn},
  J.~N. 2013, \apjl, 773, L15

\bibitem[{{Rappaport} {et~al.}(2019){Rappaport}, {Zhou}, {Vanderburg}, {Mann},
  {Kristiansen}, {Ol{\'a}h}, {Jacobs}, {Newton}, {Omohundro}, {LaCourse},
  {Schwengeler}, {Terentev}, {Latham}, {Bieryla}, {Soares-Furtado}, {Bouma},
  {Ireland }, \& {Irwin}}]{Rappaport_2019_1}
{Rappaport}, S., {Zhou}, G., {Vanderburg}, A., {et~al.} 2019, \mnras, 485, 2681

\bibitem[{{Redfield} {et~al.}(2017){Redfield}, {Farihi}, {Cauley}, {Parsons},
  {G{\"a}nsicke}, \& {Duvvuri}}]{Redfield_2017_1}
{Redfield}, S., {Farihi}, J., {Cauley}, P.~W., {et~al.} 2017, \apj, 839, 42

\bibitem[{{Roche}(1849)}]{Roche_1849}
{Roche}, E. 1849, Acad\'emie des Sciences de Montpellier: M\'emoires de la
  Section des Sciences, 1, 243

\bibitem[{{Schaefer} {et~al.}(2018){Schaefer}, {Bentley}, {Boyajian}, {Coker},
  {Dvorak}, {Dubois}, {Erdelyi}, {Ellis}, {Graham}, {Harris}, {Hall}, {James},
  {Johnston}, {Kennedy}, {Logie}, {Nugent}, {Oksanen}, {Ott}, {Rau},
  {Vanaverbeke}, {van Lieshout}, \& {Wyatt}}]{Schaefer_2018_1}
{Schaefer}, B.~E., {Bentley}, R.~O., {Boyajian}, T.~S., {et~al.} 2018, \mnras,
  481, 2235

\bibitem[{{Schlafly} \& {Finkbeiner}(2011)}]{Schlafly2011}
{Schlafly}, E.~F., \& {Finkbeiner}, D.~P. 2011, \apj, 737, 103

\bibitem[{{Smallwood} {et~al.}(2018){Smallwood}, {Martin}, {Livio}, \&
  {Lubow}}]{Smallwood_2018_1}
{Smallwood}, J.~L., {Martin}, R.~G., {Livio}, M., \& {Lubow}, S.~H. 2018,
  \mnras, 480, 57

\bibitem[{{Soker} {et~al.}(1984){Soker}, {Livio}, \& {Harpaz}}]{Soker_1984_1}
{Soker}, N., {Livio}, M., \& {Harpaz}, A. 1984, \mnras, 210, 189

\bibitem[{{Thompson} \& {Mullally}(2013)}]{Thompson2013}
{Thompson}, S., \& {Mullally}, F. 2013, Astrophysics Source Code Library,
  ascl:1304.004

\bibitem[{{Thompson} {et~al.}(2010){Thompson}, {Montgomery}, {von Hippel},
  {Nitta}, {Dalessio}, {Provencal}, {Strickland }, {Holtzman}, {Mukadam},
  {Sullivan}, {Nagel}, {Koziel-Wierzbowska}, {Kundera}, {Zola}, {Winiarski},
  {Drozdz}, {Kuligowska}, {Ogloza}, {Bogn{\'a}r}, {Handler}, {Kanaan},
  {Ribeira}, {Rosen}, {Reichart}, {Haislip}, {Barlow}, {Dunlap}, {Ivarsen},
  {LaCluyze}, \& {Mullally}}]{Thompson_2010_1}
{Thompson}, S.~E., {Montgomery}, M.~H., {von Hippel}, T., {et~al.} 2010, \apj,
  714, 296

\bibitem[{{Tremblay} {et~al.}(2011){Tremblay}, {Bergeron}, \&
  {Gianninas}}]{Tremblay2011}
{Tremblay}, P.~E., {Bergeron}, P., \& {Gianninas}, A. 2011, \apj, 730, 128

\bibitem[{{Tremblay} {et~al.}(2013){Tremblay}, {Ludwig}, {Steffen}, \&
  {Freytag}}]{Tremblay2013}
{Tremblay}, P.~E., {Ludwig}, H.~G., {Steffen}, M., \& {Freytag}, B. 2013, \aap,
  559, A104

\bibitem[{{Vanderburg} {et~al.}(2015){Vanderburg}, {Johnson}, {Rappaport},
  {Bieryla}, {Irwin}, {Lewis}, {Kipping}, {Brown}, {Dufour}, \&
  {Ciardi}}]{Vanderburg2015}
{Vanderburg}, A., {Johnson}, J.~A., {Rappaport}, S., {et~al.} 2015, \nat, 526,
  546

\bibitem[{{Veras}(2016)}]{Veras_2016_1}
{Veras}, D. 2016, Royal Society Open Science, 3, 150571

\bibitem[{{Veras} \& {G{\"a}nsicke}(2015)}]{Veras_2015_2}
{Veras}, D., \& {G{\"a}nsicke}, B.~T. 2015, \mnras, 447, 1049

\bibitem[{{Veras} {et~al.}(2014){Veras}, {Leinhardt}, {Bonsor}, \&
  {G{\"a}nsicke}}]{Veras2014}
{Veras}, D., {Leinhardt}, Z.~M., {Bonsor}, A., \& {G{\"a}nsicke}, B.~T. 2014,
  \mnras, 445, 2244

\bibitem[{{Veras} {et~al.}(2015){Veras}, {Leinhardt}, {Eggl}, \&
  {G{\"a}nsicke}}]{Veras_2015_1}
{Veras}, D., {Leinhardt}, Z.~M., {Eggl}, S., \& {G{\"a}nsicke}, B.~T. 2015,
  \mnras, 451, 3453

\bibitem[{{Veras} {et~al.}(2020){Veras}, {McDonald}, \&
  {Makarov}}]{Veras_2020_1}
{Veras}, D., {McDonald}, C.~H., \& {Makarov}, V.~V. 2020, \mnras, 492, 5291

\bibitem[{{von Hippel} \& {Thompson}(2007)}]{vonHippel_2007_1}
{von Hippel}, T., \& {Thompson}, S.~E. 2007, \apj, 661, 477

\bibitem[{{Williams} {et~al.}(2009){Williams}, {Bolte}, \&
  {Koester}}]{Williams_2009_1}
{Williams}, K.~A., {Bolte}, M., \& {Koester}, D. 2009, \apj, 693, 355

\bibitem[{{Wilson} {et~al.}(2019){Wilson}, {Farihi}, {G{\"a}nsicke}, \&
  {Swan}}]{Wilson_2019_1}
{Wilson}, T.~G., {Farihi}, J., {G{\"a}nsicke}, B.~T., \& {Swan}, A. 2019,
  \mnras, 487, 133

\bibitem[{{Zorotovic} {et~al.}(2014){Zorotovic}, {Schreiber}, \&
  {Parsons}}]{Zorotovic_2014_1}
{Zorotovic}, M., {Schreiber}, M.~R., \& {Parsons}, S.~G. 2014, \aap, 568, L9

\bibitem[{{Zuckerman} {et~al.}(2010){Zuckerman}, {Melis}, {Klein}, {Koester},
  \& {Jura}}]{Zuckerman2010}
{Zuckerman}, B., {Melis}, C., {Klein}, B., {Koester}, D., \& {Jura}, M. 2010,
  \apj, 722, 725

\end{thebibliography}

\end{document}